\documentclass[10pt,a4paper,authoryear,twocolumn]{article} 
\usepackage[utf8]{inputenc}
\usepackage[english]{babel}
\usepackage{amsmath,amssymb}
\usepackage{graphicx}
\usepackage{geometry}
\usepackage{rotating}
\usepackage{color}
\usepackage{authblk}
\usepackage{breakcites} 
\usepackage{multicol}

\usepackage{oplotsymbl} 
\usepackage{fdsymbol} 

\title{A minimal electrical model of the human heart}
\author{Jo\~ao Ol\'{\i}via,   Rui Dil\~ao}

\affil{University of Lisbon, Instituto Superior T\'ecnico, Dep. of Physics\\
Av. Rovisco Pais, 1049-001 Lisbon, Portugal}

\date{\today} 

\begin{document}

\twocolumn[
\begin{@twocolumnfalse}
 \maketitle

\begin{abstract}
We develop a minimal whole-heart model that describes cardiac electrical conduction and simulate a basic three-lead electrocardiogram (ECG). We compare our 3-lead ECG model with clinical data from a Norwegian athlete database. The results demonstrate a strong correlation with the ECGs recorded for these athletes. We simulate various pathologies of the heart's electrical conduction system, including ventricular tachycardia, atrioventricular nodal reentrant tachycardia, accessory pathways, and ischaemia-related arrhythmias, showing that the 3-lead ECGs align with the clinical data. This minimal model serves as a computationally efficient digital twin of the heart.
\end{abstract}

\smallskip

\bigskip
 \end{@twocolumnfalse}
]

\section{Introduction}

Whole-heart modelling has advanced significantly in recent years 
\cite{Qia, Bue}. Efforts in action potential modelling, combined with advances in optical imaging, have enabled researchers to explore the creation of in silico versions of human hearts, including those of specific patients, using them as predictive and diagnostic tools. Initially, in silico heart models may not aim to replicate a specific heart but rather to construct a generic one, thereby reducing the need for animal experimentation 
\cite{Tra}. Conversely, precision medicine—where therapies are custom-designed for each patient—relies heavily on the ability to develop digital twins, making the data from a specific patient critically important. By integrating the wealth of information obtained from various examinations, such as echocardiography, magnetic resonance imaging, ECG, genetic testing, and blood pressure measurements, along with knowledge of physics and physiology, it becomes possible to create computationally robust models that can replace traditional empirical diagnostic tools 
\cite{Gil}. These models provide quantitative insights into the underlying causes of specific symptoms. Numerous research groups have initiated and are currently pursuing such projects. For instance, the work carried out in the Trayanova lab employs imaging tools to create a digital twin, which is then used to predict the optimal ablation sites for patients affected by Arrhythmogenic Right Ventricular Cardiomyopathy 
\cite{Nie}. The research conducted by Viola and colleagues demonstrates the computational power required for these approaches by leveraging the processing capabilities of GPUs in the development of digital twins 
\cite{Vio}. A review by 
\cite{Tha} highlights the significant potential of digital twin development, along with the challenges associated with its implementation in clinical environments.

Several models have been developed to describe the propagation of action potential signals along cardiac tissues. The first model was introduced in 1962 by Denis Noble as a modified version of the Hodgkin-Huxley neuronal model \cite{Nob}. Additional models depicting action potentials in various heart cells have been published for both human and animal hearts \cite{Iri, Cou,Tus}. Over the years, more detailed models have emerged, the vast majority of which are available at the CellML model repository (https://www.cellml.org). These models are typically categorised into three generations, each with increasing complexity. The first-generation models, exemplified by the Beeler-Reuter model, include a limited number of ion channels such as Na$^+$, K$^+$, and Ca$^{2+}$ and consist of fewer than 10 differential equations \cite{Bee}. The inclusion of Ca$^{2+}$ was a crucial milestone, as this ion is primarily responsible for the electromechanical coupling essential to the heart’s function. The second-generation models increase the number of ion channels from 10 to 20 and the differential equations from 20 to 40. The third-generation models currently under development are significantly more complex, taking into account specific biological processes and dynamics. Although the trend has been towards increasingly complex models \cite{Qua, Fed}, simplified variants still yield significant results, as seen in the ventricular action potential developed by \cite{Bue}.  

In this paper, we develop a minimal whole-heart model that describes the basic anatomy of the heart with minimal details to ensure computational efficiency. In section~\ref{sec:geom}, we model the four heart cavities as four two-dimensional square domains with appropriate boundary conditions. Each square domain contains a lattice of points representing heart cells (myocytes) that communicate with their neighbouring cells. Anatomical structures such as valves and nodal points are included to facilitate the analysis of connectivity and electrical conduction through the heart. In section~\ref{sec:HH2D}, we introduce a simplified two-dimensional Hodgkin-Huxley type model that characterises the electrical states of the human heart as they propagate through the cell lattices. In section~\ref{sec: Method}, we simulate the complete cycle of a healthy heart, demonstrating that the fundamental electrophysiological properties are present in the model. In section~\ref{sec:WHM}, we reconstruct the first three leads of an ECG, comparing it with clinical data. In section~\ref{sec:diseases}, we simulate several heart pathologies, including ventricular tachycardia, atrioventricular nodal reentrant tachycardia, accessory pathways, and ischaemia-related arrhythmias. Finally, in section~\ref{sec:concl}, we summarise the main results of the paper.

 \section{The anatomical geometry of the human heart}\label{sec:geom}

One of the most important concepts underlying this work is the spatial modelling of the heart. For each approximately spherical heart cavity, a two-dimensional square domain with the appropriate boundary conditions will be employed, representing the surface of a sphere. Each surface will depict the muscular tissue with cells arranged in a square lattice. Each point of the square lattice represents a cell characterised by different parameter values and electrical states.

\begin{figure}
    \centering
    \includegraphics[width=0.5\textwidth]{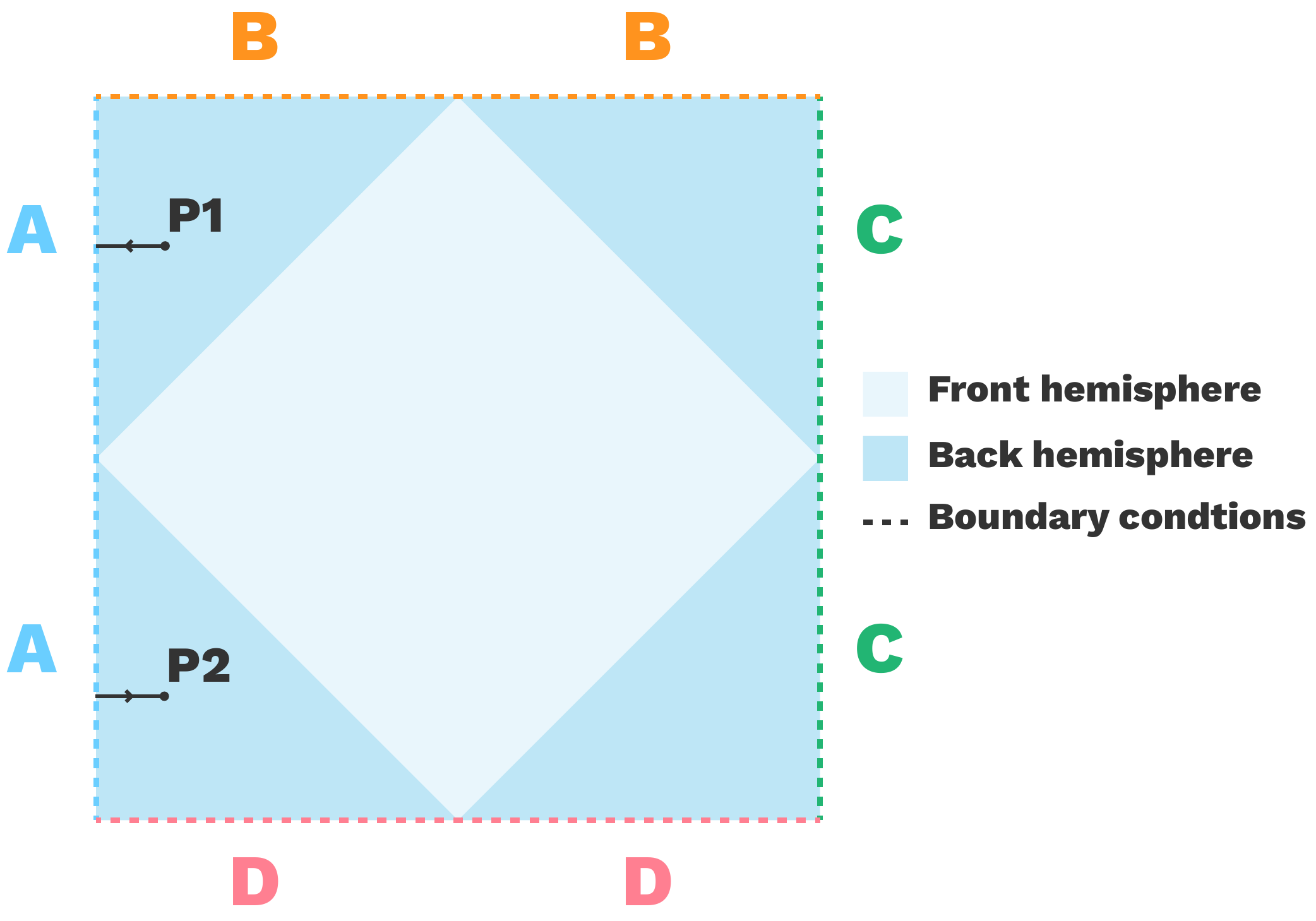}
    \caption{A 2D map (chart) of a spherical surface is used to simulate the heart cavities. The half-sides displaying the same letter are identified. The front and back hemispheres of the spherical surface are differentiated by lighter and darker shades of blue, respectively.}
    \label{fig:map1}
\end{figure}

In figure~\ref{fig:map1}, the projection of the spherical surface is depicted. Each side of the square is labelled with a letter from A to D and is divided into two halves. The lattice points in the halves sharing the same letter are identified. This results in a two-dimensional surface that is homeomorphic to a two-dimensional sphere. This construction represents a chart of the 2-sphere \cite[Appendix A.1.2]{Dil}.

The four heart cavities are modelled by four identical squares discretised into $76 \times 76$  lattice cells. The number of cells has been chosen to ensure that the action potential has an average propagation time of $0.5$~m/s for the selected integration time step $\Delta t $. With this geometry serving as the fundamental component of our heart model, we now outline each of the heart's cavities along with their respective anatomical structures and connections.

The coordinate system used to locate each heart cell or myocyte is $P_{\hbox{\footnotesize Structure}}=(x,y)_{\hbox{\footnotesize Cavity}}$, where $(x,y)$ denotes the two-dimensional integer coordinates of the lattice points in the heart cavities: Right Atrium (RA), Left Atrium (LA), Right Ventricle (RV), and Left Ventricle (LV). The origins of the coordinates are at the top-left vertices of each square lattice. 

Heart cavities are connected by linear chains or bundles of cells and are referred to as $P_{\hbox{\footnotesize bundle}}=(x)_{\hbox{\footnotesize bundle}}$, where $x$ is the number of cells in the chain.

\begin{figure*}[t!]
    \centering
    \includegraphics[width=.80\textwidth]{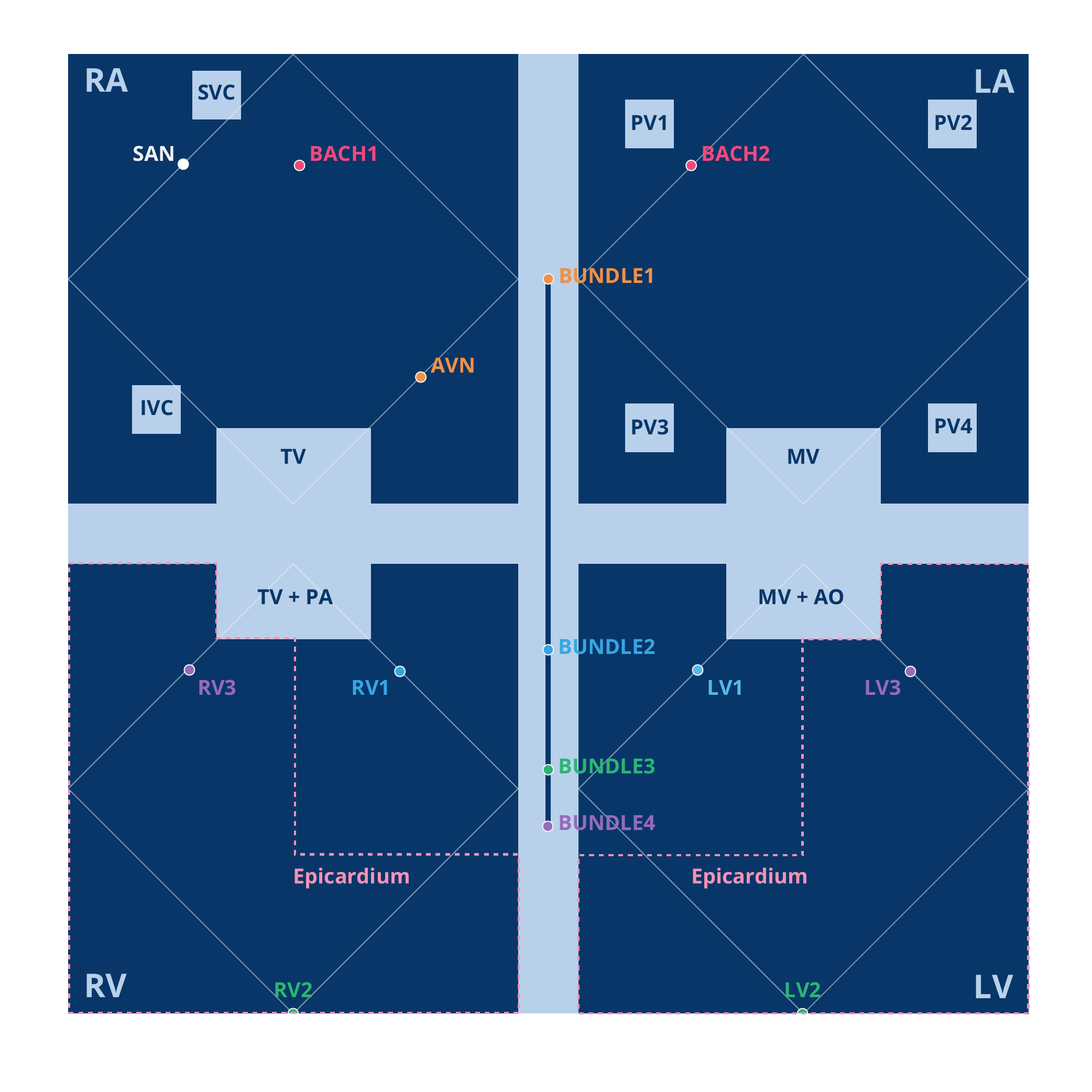}
    \caption{2D map of the entire heart includes its four cavities: the right and left atria and ventricles. BACH1 and BACH2 serve as the connection points for the two atria, simulating Bachmann's bundle, which is reduced to a point in this model. The bundle line links the atria to the ventricles. The Atrioventricular Node (AVN) connects to the bundle points BUNDLE1. The bundle points BUNDLE2, BUNDLE3, and BUNDLE4 are connected to RV1/LV1, RV2/LV2, and RV3/LV3, respectively, simulating the His and Purkinje fibres. The dashed lines indicate the boundaries of the ventricular epicardial layers, added to account for the thickness of the ventricular walls.}
    \label{fig:WH_map}
\end{figure*}

In figure~\ref{fig:WH_map}, the final geometry of the entire heart is displayed. The diamond-shaped lines serve as references for the front and back hemispheres, akin to what was presented in figure\ref{fig:map1}. 

The RA is characterised by the presence of three zero-flux anatomical structures. These structures comprise the Superior and Inferior Vena Cavae (SVC and IVC), represented by squares with a side length of $a = 8$. Their fixed central positions are $P_{\hbox{\footnotesize SVC}} = (25,8)_{\hbox{\footnotesize RA}}$ and $P_{\hbox{\footnotesize IVC}} = (15,60)_{\hbox{\footnotesize RA}}$, along with the Tricuspid Valve (TV), modelled as a rectangle with dimensions $13 \times 26$, situated at the bottom centre of the square. The boundaries of these three anatomical structures are characterised by zero-flux boundary conditions.
The Sinoatrial Node (SAN) is located at $P_{\hbox{\footnotesize SAN}} = (19,19)_{\hbox{\footnotesize RA}}$. The BACH1 and BACH2 points connect the right to the left atria, simulating Bachmann's Bundle, with $P_{\hbox{\footnotesize BACH1}}=(30,25)_{\hbox{\footnotesize RA}}$ and $P_{\hbox{\footnotesize BACH2}} = (19,19)_{\hbox{\footnotesize LA}}$. The electrical states of BACH1 and BACH2 are the same. The Atrioventricular Node (AVN) with coordinates $P_{\hbox{\footnotesize AVN}} = (57,57)_{\hbox{\footnotesize RA}}$ connects to the first point of BUNDLE1, responsible for the linear connection between the atria and ventricles, simulating the His bundle.

 Within the LA, the Pulmonary Veins (PV) are represented as zero-flux squares, each with a side length of $a = 8$, situated at $P_{\hbox{\footnotesize PV1}} = (13,13)_{\hbox{\footnotesize LA}}$, $P_{\hbox{\footnotesize PV2}} = (63,13)_{\hbox{\footnotesize LA}}$, $P_{\hbox{\footnotesize PV3}} = (13,63)_{\hbox{\footnotesize LA}}$, and $P_{\hbox{\footnotesize PV4}} = (63,63)_{\hbox{\footnotesize LA}}$. The Mitral Valve (MV) is located at the bottom centre of the LA.

The connection between the atria and ventricles in the human heart occurs through three primary structures: the His bundle, the bundle branches, and the Purkinje fibres. For simplicity, we have described these structures using a single channel, referring to it as \textit{bundle}. This structure varies in length depending on the heart rate. 

Within the bundle, $P_{\hbox{\footnotesize BUNDLE1}} = (1)_{\hbox{\footnotesize bundle}}$ connects directly to the AVN, while $P_{\hbox{\footnotesize BUNDLE2}} = (60)_{\hbox{\footnotesize bundle}}$, $P_{\hbox{\footnotesize BUNDLE3}} = (80)_{\hbox{\footnotesize bundle}}$, and $P_{\hbox{\footnotesize BUNDLE4}} = (90)_{\hbox{\footnotesize bundle}}$ are linked to the ventricles RV1/LV1, RV2/LV2, and RV3/LV3, respectively. These points were specifically selected to model the path taken by the Purkinje fibres across the ventricular tissue. The points $P_{\hbox{\footnotesize RV1}} = (57,19)_{\hbox{\footnotesize RV}}$ and $P_{\hbox{\footnotesize LV1}} = (19,19)_{\hbox{\footnotesize LV}}$ are situated at the interventricular septum, which is the first region excited by the Purkinje fibres. The points $P_{\hbox{\footnotesize RV2}} = (38,76)_{\hbox{\footnotesize RV}}$ and $P_{\hbox{\footnotesize LV2}}= (38,76)_{\hbox{\footnotesize LV}}$ denote the apex of the ventricles, while $P_{\hbox{\footnotesize RV3}} = (19,19)_{\hbox{\footnotesize RV}}$ and $P_{\hbox{\footnotesize LV3}} = (57,19)_{\hbox{\footnotesize LV}}$ signify the base of the ventricles. 

Similar to the atria, the ventricles have non-conducting regions that serve as the valves; however, in this case, they also represent the obstruction caused by the Pulmonary Artery (PA) and the Aorta (AO) for the right and left ventricles, respectively. 

In addition to the basic structure of the heart described earlier, a ventricular epicardial layer has been incorporated into the aforementioned ventricular layers to ensure the correct direction of ventricular repolarisation, which will be analysed in the next section. In human ventricles, the outer layer of the muscular walls, the epicardium, exhibits shorter action potentials compared to the myocardium and the endocardium. Consequently, the last cells to depolarise are the first to repolarise, resulting in the repolarisation wavefront being opposite to the depolarisation wavefront. The boundaries of these epicardial layers are delineated across the ventricles with orange dashed lines. They do not form complete spheres, as they do not encircle the ventricles in the interventricular septum region. Each ventricular layer is vertically connected to its corresponding epicardial layer; that is, each cell on one layer is linked to the cell with the same coordinates on the other layer.

\section{Propagation of the electrophysiological stimulus}
\label{sec:HH2D}

The equations that describe how the electric stimulus propagates through cardiac cells are an adapted Hodgkin-Huxley (HH) reaction-diffusion model in two spatial dimensions, controlled exclusively by a potassium-sodium gating variable \cite{Bra}. The choice of this adapted version of the model over the original four-dimensional HH model is based on the overall intention of simplicity. The model's reaction-diffusion partial differential equations are
\begin{equation}
\left\{\begin{array}{rcl}\displaystyle
    C_{m} \frac{dV}{dt} &=&\displaystyle  I+D\left(\frac{\partial^2V}{\partial x^2} +\frac{\partial^2V}{\partial y^2}\right)-g_K n^4(V-V_K)\\ [10pt]\displaystyle
    &&\displaystyle -g_{Na}m_{\infty}^{3}(c(I)-n)(V-V_{Na})\\  \displaystyle
    \frac{dn}{dt} &=&  \alpha_{n}(V)(1-n)-\beta_{n}(V),
    \end{array}\right.
    \label{eqn:HH2D}
\end{equation}
where
$$\begin{array}{rcl}
m_{\infty} &=&\displaystyle \frac{\alpha_{m}(V)}{\alpha_{m}(V)+\beta_{m}(V)}\\ [8pt]\displaystyle
    \alpha_{m}(V) &=&\displaystyle 0.1\phi\frac{V+25}{e^{(V+25)/10}-1} \\ [8pt]\displaystyle
    \beta_{m}(V) &=&\displaystyle 4\phi e^{V/18}  \\ [8pt]\displaystyle
    \alpha_{n}(V) &=&\displaystyle 0.01\phi\frac{V+10}{e^{(V+10)/10}-1} \\ [8pt]\displaystyle
    \beta_{n}(V) &=&\displaystyle 0.125\phi e^{V/80}\\ [8pt]\displaystyle
    c(I)&=& \begin{cases}
      1.0 & \hbox{for} \, I \in [0,2)\\
      1.046I^{-0.077} & \hbox{for} \, I \in [2,160],
    \end{cases}
    \end{array}
$$
and $n$ represents the potassium-sodium gating variable. We selected the HH value $C_m = 50$~$\mu$F/cm$^2$ for the transmembrane capacitance. The Nernst potentials for both sodium and potassium channels are $V_{Na} = -115$~mV and $V_{K} = -12$~mV. The conductances for sodium and potassium channels are $g_{Na} = -120 $~mS/cm$^2$ and $g_K = 36$~mS/cm$^2$, respectively, with $\phi = 1$. For a model with more gating variables, see \cite{Alo}.

\begin{table*}
\centering
\caption{Initial conditions and parameters utilised in the simulation of the heartbeat of a healthy heart with a frequency of 60 bpm.}
{\footnotesize
\begin{tabular}{cccccc} \hline
    {cells} & {$V(t = 0)$ (mV)} &{$n(t = 0)$}&{$I$ ($\mu $A/cm$^2$)}  & {$g_{K}$ (mS/cm$^2$)}  & {$R$ (k$\Omega $cm$^2$)}\\ \hline
    $P_{\hbox{SAN}} =(19,19)_{RA}$& -10.95&0.17& 2.95   & 36  &0.00675 \\
    Ventricle cells&  -10.95&0.17& 0       & 16.54   &0.00675 \\
    All other cells&  -10.95&0.17& 0       & 36   &0.00675\\ \hline
\end{tabular}
}
\label{tab:healthy60bpm}
\end{table*}

For myocytes, the current in equation \eqref{eqn:HH2D} is set to $I = 0$~$\mu$A/cm$^2$. The oscillation period of the SAN relates to the input current $I$ through the expression $T(I_{SAN}) = 1.7I_{SAN}^{-0.45}$~s, such that the standard heartbeat of $60$~bpm ($T = 1$~s) is achieved by setting $I_{SAN} = 2.95$~$\mu$ A/cm$^2$ (\cite{Bra}). 

The equations    \eqref{eqn:HH2D} were discretised and numerically solved using the forward-Euler method. The diffusion coefficients, which describe the propagation along the myocytes, were calibrated according to the relation $D \Delta t/\Delta x^2=1/6$, where $\Delta t $ is the integration time step and $\Delta x$ is the spatial scale of the cavity point lattices. Additionally, $D=\Delta x^2/R$, where $R$ is the resistivity parameter characterising the heart tissue (\cite{Dil2} and \cite{Bra}).

To ensure that the ventricular AP is longer than the epicardial AP, the potassium channel conductance $g_K$ in the ventricular layer cells is reduced to create an extended plateau phase. This duration is influenced by heart rate; the faster the oscillation frequency of the sinoatrial node (SAN), the smaller the difference between the AP durations of both layers. The variation in the interval between depolarisation and repolarisation of the ventricles as a function of heart rate has been well studied. Numerous expressions have been proposed over the years to adjust the QT ECG interval for various heart rates, a concept known as the QTc (corrected QT interval). Here, we have employed the expression 
$$
    \label{eq:gkcorr}
    g_{Kcorr} = \frac{1}{-0.018}\left(0.42\left(\sqrt[3]{1.7I_{SAN}^{-0.45}}\right)-0.724\right)
$$
adapted from \cite{Fri} and \cite{Bra}. For a heart rate of 60 bpm, the conductances of potassium channels in the ventricular layer cells are set at $g_{Kcorr} = 16.54$~mS/cm$^2$, while all other cells, including those in the epicardial layer, retain the standard value of $g_{K} = 36$~mS/cm$^2$.
 
Concerning the propagation of the AP along the bundle, we have assumed that the length of the bundle is
$$
    \label{eq:lbundle}
    l_{bundle} = 0.7\left(-0.69\left(\frac{60}{1.7I_{SAN}^{-0.45}}-60\right)+85.7\right)+30
$$
rounded to the nearest integer. This is based on the observation that, in a control group, an increase of $6.9$~ms was noted in the interval between atrial and ventricular depolarisation for every 10-beat increase in heart rate, thus indicating a linear relationship between the two \cite{Att}.

\section{Dynamics of the whole heart}
\label{sec: Method}

The temporal evolution of the whole heart's electrophysiological state has been simulated in Python and is available on GitHub \cite{GH}. In all the simulations presented in this paper, we have chosen $\Delta t=0.375$~ms for the numerical integration of equations \eqref{eqn:HH2D}. The spatial scale is $\Delta x=0.93$~mm to accommodate heart dimensions. Other parameters for a healthy heart with a heart rate of 60 bpm are shown in Table \ref{tab:healthy60bpm}.

\begin{sidewaysfigure*}
    \centering
    \includegraphics[width=0.99\textwidth]{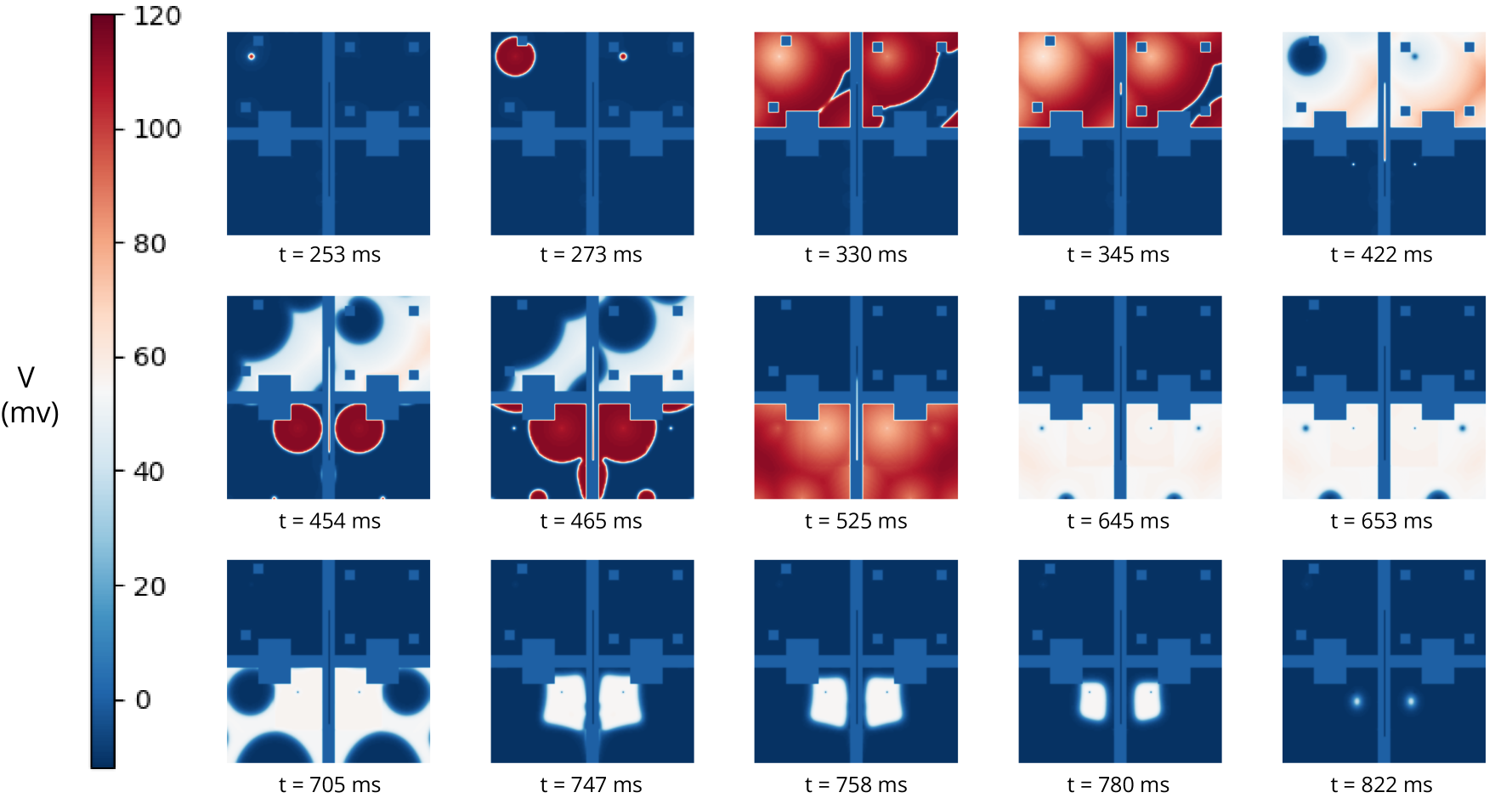}
    \caption{Propagation of the membrane potential across the four heart chambers, calculated using the model equations \eqref{eqn:HH2D}, during a single heartbeat at a heart rate of $60$~bpm. The AP signal originates at the SAN at time $t=253$ ms. Each frame provides a complete view of the heart.}
    \label{fig:snap1}
\end{sidewaysfigure*}

In figure~\ref{fig:snap1}, we present 15 frames from a complete heartbeat, carefully selected to illustrate the critical stages of each cycle. This heartbeat begins at $t = 253$~ms, when the stimulus from the SAN propagates to the adjacent cells. At $t = 273$~ms, the signal reaches the left atrium through the connection from the BACH1 to the BACH2 point. The signal then arrives at the AVN and spreads to the BUNDLE1 point at $t = 330$~ms. As the excitation continues to advance through the bundle, the atria begin to repolarise, depicted by the progression of the blue wavefront from $t = 422$~ms to $t = 465$~ms. At $t = 454$~ms, the signal reaches the first contact points with the ventricles, the RV1 and LV1 points, initiating the ventricles' depolarisation. As the signal reaches the tenth from last cell of the bundle, at $t = 454$~ms, the RV2 and LV2 points depolarise, propagating the signal from the apex of the ventricles. At $t = 465$~ms, the signal arrives at the final contact point between the bundle and the ventricles, and the excitation emanates through the RV3/LV3 points. From just after $t = 525$~ms to just before $t = 645$~ms, the ventricular myocytes are in the extended plateau phase made possible by the differing potassium conductance of $g_{K} = 16.5$~mS/cm$^2$. At $t = 645$~ms, the epicardium, which is not displayed but is understood to be atop the ventricular layers, commences repolarisation at the apex and then at the base, propagating this depolarisation to the RV2/LV2 and RV3/LV3 points. This repolarisation then continues in opposition to the primary depolarisation wave that originated at $t = 454$~ms until the end of the cycle at $t = 822$~ms.

\begin{figure}
	\centering
	\includegraphics[width=0.5\textwidth]{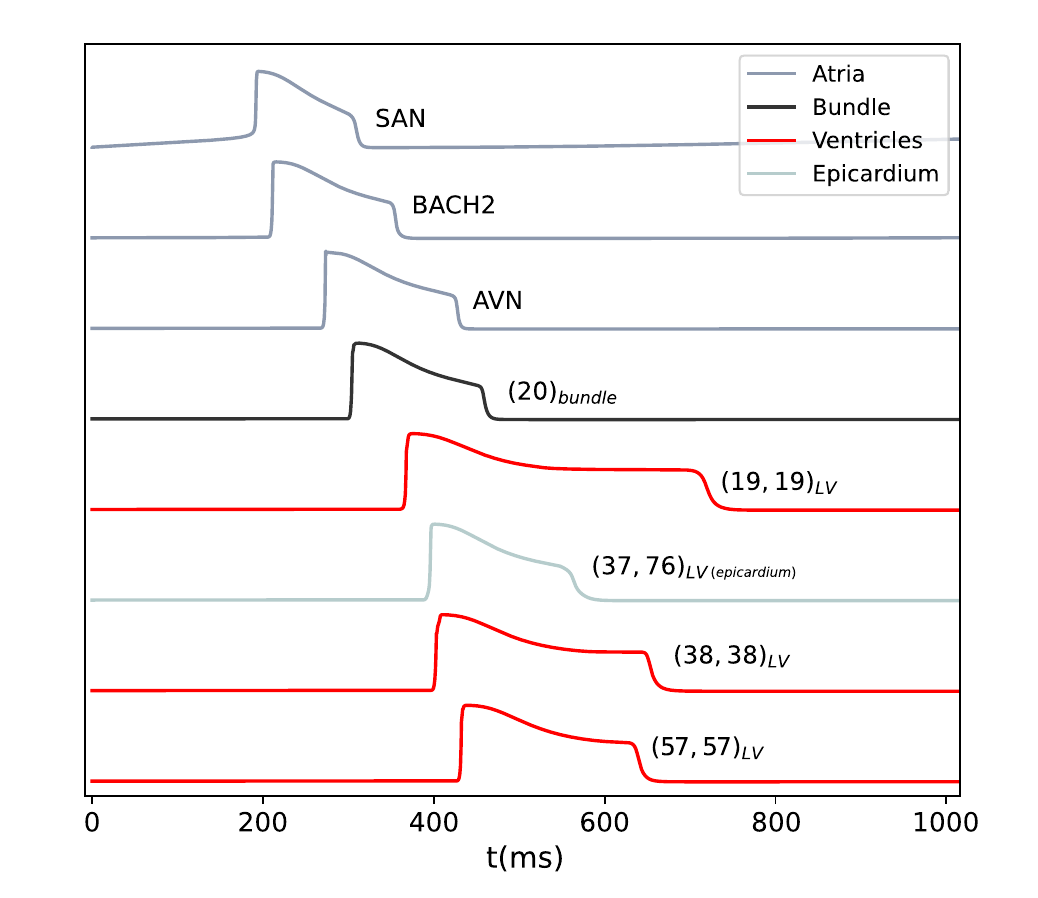}
	\caption{Action potential time profiles at selected heart cells. The darker grey corresponds to atrial cells, black corresponds to a bundle cell, red corresponds to ventricular cells, and light grey corresponds to an epicardial cell.}
	\label{fig:AP_plot}
\end{figure}

A plot of the action potentials of selected cells is presented in figure~\ref{fig:AP_plot}. Each line represents the action potential of a point of interest during each heartbeat.  The signal originates at the SAN and travels through the $BACH1-BACH2$ point to the LA. Subsequently, it arrives at the AVN, which propagates the signal to the bundle. After passing through the bundle, the signal is dispersed throughout the ventricles. The first red line indicates the cell $(19,19)_{LV}$ (the top leftmost point of the left ventricle), which is the first to excite. The cell at the centre of the frontal hemisphere, $(38,38)_{LV}$, experiences depolarisation, followed by the bottom leftmost cell at $(57,57)_{LV}$. A key characteristic of these three APs of the LV cells is their repolarisation. The later a cell is excited, the earlier it begins to repolarise, as demonstrated by the differences among these three APs. These dynamics explain why the T-wave is an upright deflection in leads I and II, a phenomenon made possible by the presence of the epicardial layers. The light blue line, representing an epicardial cell, illustrates that the AP duration in epicardial cells is similar to that of atrial myocytes, while the longest AP in the ventricles is approximately twice as long. The complete dynamics of the epicardial layer will be discussed in the following section.

\section{The Electrocardiogram} \label{sec:WHM}

The electrocardiogram serves as the benchmark by which we assess the clinical validity of our model. Electrodes placed on a patient's skin measure the local flow of ions resulting from the electrical field generated in the heart \cite{End}. The electrical impulse travels through the extra myocardial region, which consists of resistive tissue and fluid surrounding the heart within the human body, while the potential difference captured by the electrodes is reflected in an ECG. 

We measure the differences in membrane potential $V$ between pairs of neighbouring cells based on the ECG lead position. This method accounts for the average electrical field generated in the heart in each spatial direction. The expressions for the first three leads of the ECG are
\begin{equation}
	\label{eqn:leadI}
	\begin{aligned}
		&<\Delta V_{ECG}^{I}(t)> =  \\
		&-\frac{1}{N_{total}}\sum^{4}_{i = 1}\alpha_i \sum_{(x,y) \in \Omega_f} z_{(x,y)}.\left( V_{(x+1,y+1)}^{t}-V_{(x,y)}^{t} \right)
	\end{aligned}
\end{equation}
\begin{equation}
	\label{eqn:leadII}
	\begin{aligned}
		&<\Delta V_{ECG}^{II}(t)> =  \\
		&-\frac{1}{N_{total}}\sum^{4}_{i = 1}\alpha_i \sum_{(x,y) \in \Omega_f} z_{(x,y)}.\left( V_{(x,y+1)}^{t}-V_{(x,y)}^{t} \right)
	\end{aligned}
\end{equation}
\begin{equation}
	\label{eqn:leadIII}
	\begin{aligned}
		&<\Delta V_{ECG}^{III}(t)> =  \\
		&\frac{1}{N_{total}}\sum^{4}_{i = 1}\alpha_i \sum_{(x,y) \in \Omega_f} z_{(x,y)}.\left( V_{(x+1,y)}^{t}-V_{(x,y)}^{t} \right),
	\end{aligned}
\end{equation}
where
$$
		z_{(x,y)}= \sqrt{\left(\frac{l}{2}\right)^2-\left(x-\frac{l}{2}\right)^2-\left(y-\frac{l}{2}\right)^2},
$$
and the subscripts in the membrane potential $V(t)$ indicate the cell coordinates $(x,y)$. $\Omega_f$ denotes the set of all cells in the front hemisphere of each heart cavity, comprising $N_{total}$ cells, and $\alpha_i$ is a scaling coefficient unique to each heart cavity. The atria and ventricles possess different muscle masses, thus demonstrating distinct contributions to the heart's electrical signature. In this context, a simple coefficient is employed to differentiate the contributions of each cavity. The index $i$ designates each cavity, with $i = 1, 2, 3, 4$ corresponding to the RA, LA, RV, and LV, respectively. Anatomical data informed the values assigned to these coefficients, considering wall thickness and surface area. The selected values are $\alpha_1 =\alpha_2 = 1$, $\alpha_3 = 5$, and $\alpha_4 = 15$, ensuring both atria contribute similarly while the ventricles possess ten times the muscle mass of the atria. The weight parameter $z_{(x,y)}$ accounts for the curvature of the heart.  
The constant $l$ is defined as the side length of each cavity,  $l = 76$ cells. As the cardiac muscle is arranged in an approximately spherical shape around the cavities, the surfaces of the muscle walls will contribute differently to the potential difference measured in each ECG electrode.  The weight parameter  $z_{(x,y)}$ ensures that each cell's contributions to the ECG signal reflect this geometry.

There is symmetry between the front and back hemispheres, apart from the arteries and veins, which do not affect the propagation direction of the AP signal. Therefore, calculating the ECG on both sides of the cavities would be unnecessary. 

\begin{figure}
    \centering
    \includegraphics[width=0.5\textwidth]{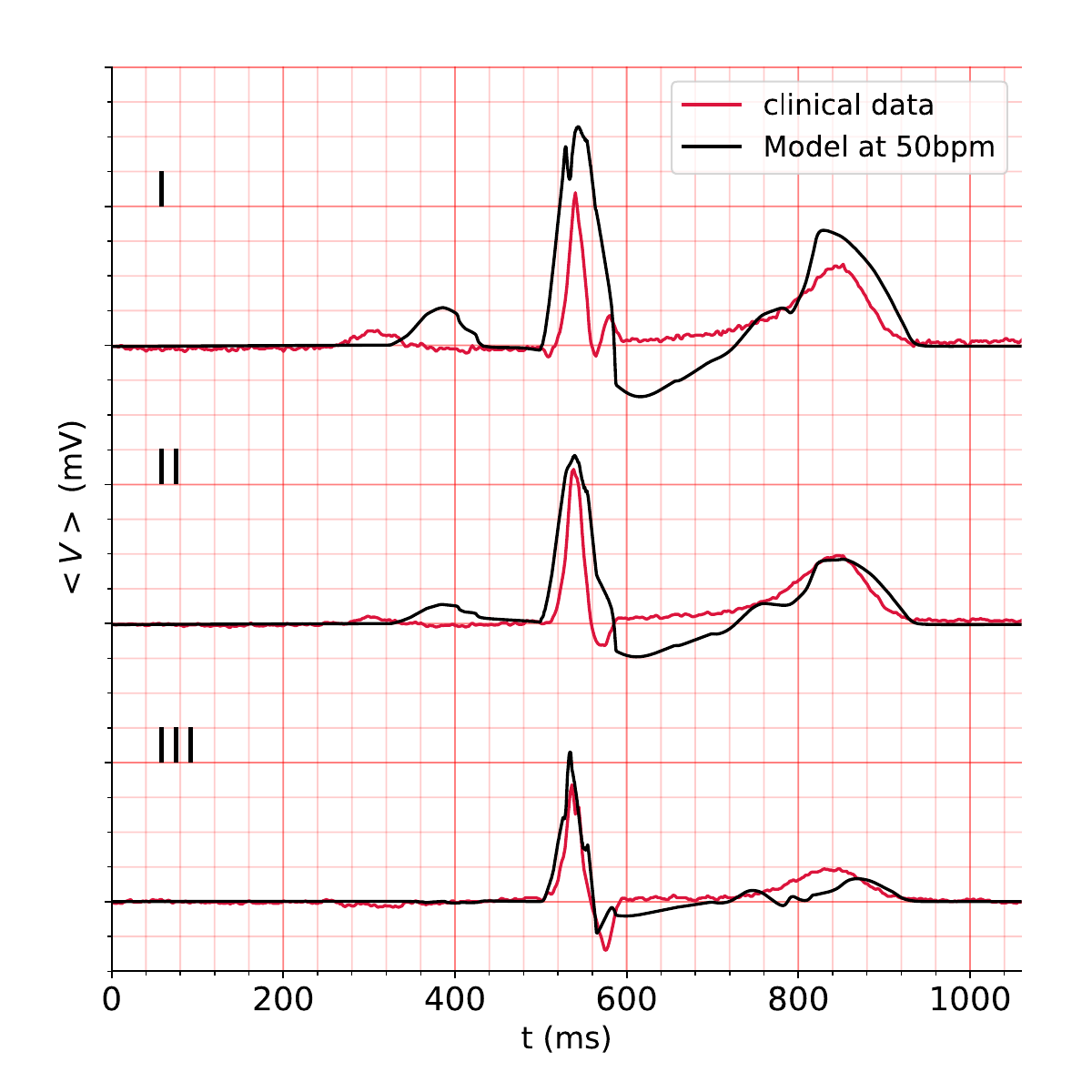}
    \caption{In red, we present the 3-lead ECG of an athlete with a heart rate of $50$~bpm. In black, we depict the model simulation obtained with $I_{SAN} = 2.2$~$\mu$A/cm$^2$. Each small square measures $0.2$~mV along the vertical direction and $40$~ms in the horizontal direction.}
    \label{fig:ECG_comp}
\end{figure}

Figure~\ref{fig:ECG_comp} illustrates the simulation results of an ECG obtained using equations \eqref{eqn:leadI}-\eqref{eqn:leadIII}, compared with the ECG of athlete number 5 from the Norwegian Endurance Athlete ECG database \cite{Sin}. A heart rate of 50 bpm was chosen for the simulated heart to match that of the athlete. 

According to the database, this athlete exhibits sinus bradycardia, sinus arrhythmia, and first-degree AV block, but otherwise has a normal ECG. We observe a high level of agreement when comparing the results from the three leads between our model and the clinical data. Both ECGs were aligned to the peaks of the QRS complex, and we notably observe deviations in the P waves. These deviations can be attributed to the first-degree AV block from which the patient suffers, characterised by PR intervals exceeding 0.2 seconds. In this case, the athlete has a PR interval of 0.26 seconds, while our model indicates a PR interval of 0.18 seconds. Additionally, subtle differences are especially evident in leads I and II, not only in the overall shape of the QRS complex but also in the form of the wave that marks the QT interval. The lack of a flatter section separating the QRS and the T-wave on our model's ECG can be explained by the shape of the action potential generated by the model equations \eqref{eqn:HH2D}. While the action potential of ventricular myocytes features a flatter plateau phase, characterised by calcium ion influx, our model shows a much steeper post-depolarisation curve, leading to changes in potential differences across the ventricles through their action potentials.

\section{Arrythmias, accessory pa\-th\-ways and tissue death}\label{sec:diseases}

\subsection{Ventricular tachycardia}

Ventricular tachycardia is a condition characterised by an increased heart rate, where the signal originates from the ventricle instead of the SAN, often due to spiral waves. One potential origin of spiral waves is the emergence of an excited region in the ventricular tissue \cite{Bea}. Such dynamics were modelled in our digital organ by establishing specific initial conditions to generate a spiral wave in the left ventricle, as shown in Table~\ref{tab:snaps_tachy}. 

\begin{table*}[]
\centering
\caption{Initial conditions and parameters for the model of the heart leading to left ventricular tachycardia.}
{\footnotesize
\begin{tabular}{cccccc} \hline
    {cells} & {$V(t = 0)$ (mV)} &{$n(t = 0)$}&{$I$ ($\mu$A/cm$^2$)}  & {$g_{K}$ (mS/cm$^2$)}  & {$R$ (k$\Omega$cm$^2$)}\\ \hline
    $(27\leq x\leq28,16\leq y\leq65)_{LV}$& 120&0.17& 0   & 36  &0.00675 \\
    $(29\leq x\leq40,16\leq y\leq65)_{LV}$& -10.95&0.99& 0   & 36  &0.00675 \\
    All other cells&  -10.95&0.17& 0       & 36   &0.00675\\ \hline
\end{tabular}
}
\label{tab:snaps_tachy}
\end{table*}

\begin{figure}
    \centering
    \includegraphics[width=0.5\textwidth]{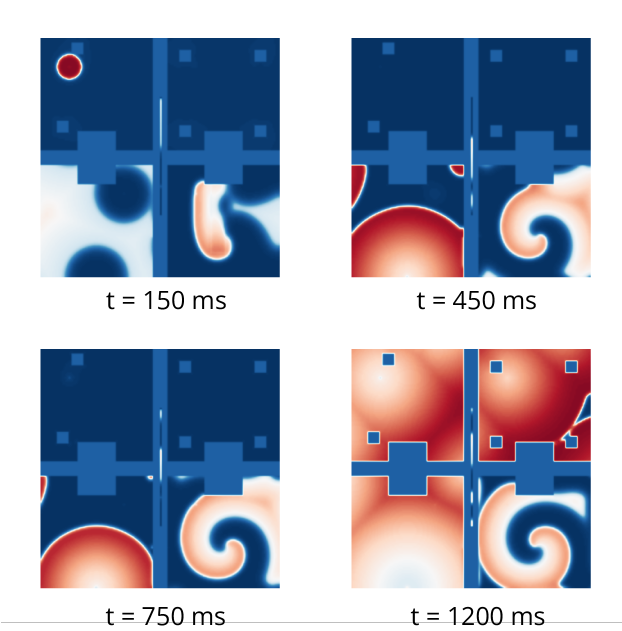} 
    \caption{Propagation of the membrane potential, where a perturbation in a rectangular region of size $14\times 50$ was introduced in the left ventricle, with the initial conditions in Table~\ref{tab:snaps_tachy}. A permanent spiral develops with a rotating arm.}
    \label{fig:snaps_tachy}
\end{figure}

Figure~\ref{fig:snaps_tachy} shows the evolution of the modelled heart with the initial conditions given in Table~\ref{tab:snaps_tachy}. A single spiral arm forms in the Left Ventricle as a result of these initial conditions. The spiral continues indefinitely, and reentry persists.

\begin{figure}[h]
    \centering
    \includegraphics[width=0.45\textwidth]{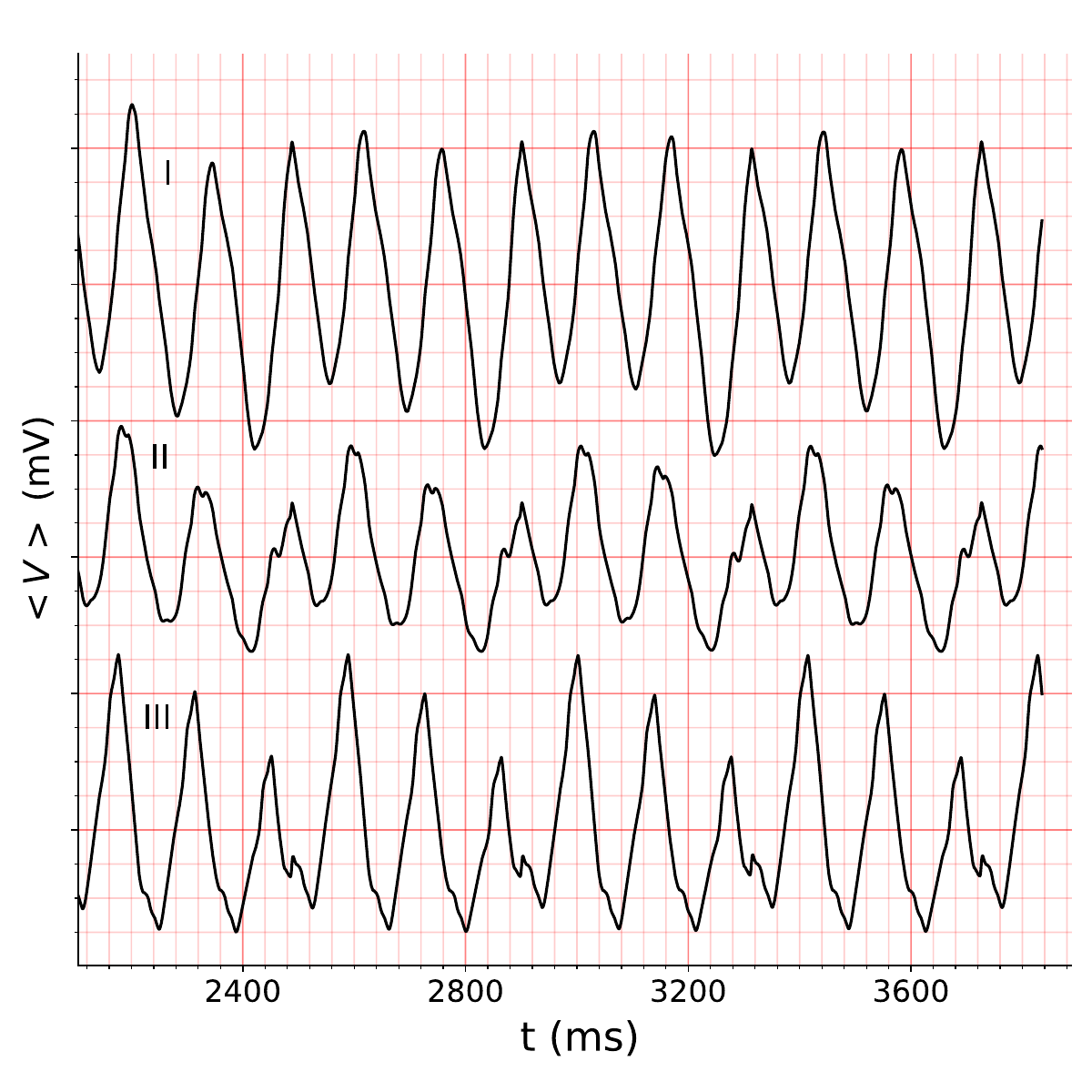} 
    \caption{Simulated the 3-lead ECG of the heart experiencing ventricular tachycardia shown in figure \ref{fig:snaps_tachy}. A non-monomorphic wave is apparent due to issues with the connection between the ventricles. The oscillation period of the ECG is approximately $T_{vtachy} = 145$~ms, contrasting with the period $T \simeq 1000$~ms of our reference healthy heart ($60$~bpm).}
    \label{fig:ecg_tachy}
\end{figure}

Displayed in figure~\ref{fig:ecg_tachy} is the modelled ECG of the heart shown in figure \ref{fig:snaps_tachy}. The shape of the healthy ECG from figure \ref{fig:ECG_comp} has been replaced with a roughly uniform oscillation. The period of oscillation is now approximately $T_{vtachy} = 145$~ms, which corresponds to a heart rate of $413$~bpm. This is extremely elevated, even for ventricular tachycardias. The wave is not entirely uniform; a period-3 pattern is evident from leads II and III, indicating the activation of the right ventricle. Such activation is not feasible for every spiral rotation due to the refractory period of the connecting cells; the right ventricle activates only after every third rotation of the spiral arm.

A limitation of our model concerns the connection between the ventricles. In this work, the connection is established at only three points rather than being continuous. In reality, the ventricles are not electrically isolated; they share the interventricular septum, meaning they are not two separate surfaces like the atria, which connect solely via Bachmann's bundle. Correcting the connections between the ventricles will remain a task for future work. If this imprecision were not present, we would expect the ECG to show monomorphic tachycardia, where each oscillation has very similar dynamics. However, under these conditions, the three leads are especially alike. Due to the rotational nature of the spiral wave, there is no preferred direction for the signal in the left ventricle, which mainly influences the ECG signature given its size.
The amplitude of the waves is also reduced due to the ineffective action of the ventricles.

\subsection{Atrioventricular nodal reentry tachycardia}

The AVN is not merely a passage between the right atrium and the bundle of His, as modelled for a healthy heart. Rather, as its name suggests, this node is a cluster of cells with specific dynamics. In certain individuals, the AVN possesses two pathways through which the signal can travel: a slower one and a faster one  \cite{Geo}. The cells in the fast pathway have longer refractory periods than those in the slow branch. This, combined with premature excitation in the right atrium, disrupts the symmetry between the two pathways, provided that the premature excitation occurs early enough for the signal to reach the fast pathway while its cells are still recovering from the previous excitation. Consequently, under such conditions, the signal will only travel through the slow path, while no signal will pass through the faster one, resulting in re-entry in a roundabout effect. New wavefronts no longer originate at the SAN but at the AVN. The PR interval is significantly shorter, and the T-wave and P-wave become overlaid. 

\begin{table*}[]
\centering
\caption{Initial conditions and parameters for the simulated heart affected by atrioventricular nodal reentry Tachycardia.}
{\footnotesize
\begin{tabular}{cccccc} \hline
    {cells} & {$V(t = 0)$ (mV)} &{$n(t = 0)$}&{$I$ ($\mu$A/cm$^2$)}  & {$g_{K}$ (mS/cm$^2$)}  & {$R$ (k$\Omega$cm$^2$)}\\ \hline
    $(25,2\leq y\leq23)_{AVN}$& -10.95&0.17& 0   & 36  &0.223 \\
    $(6,2)_{AVN}$&  -10.95&0.17& 0       & 36   &0.00675(right),$\infty$ (left)\\
    $P_{\hbox{SAN}}=(19,19)_{RA}$& -10.95&0.17& 2.95   & 36  &0.00675 \\
    Ventricle cells&  -10.95&0.17& 0       & 16.54   &0.00675 \\
    All other cells&  -10.95&0.17& 0       & 36   &0.00675\\ \hline
\end{tabular}
}
\label{tab:avn_snaps}
\end{table*}

\begin{table*}[]
\centering
\caption{Initial conditions and parameters for the model heart affected by Wolff-Parkinson-White syndrome.}
{\footnotesize
\begin{tabular}{ccccccc} \hline
    {cells} & {$V(t = 0)$ (mV)} &{$n(t = 0)$}&{$I$ ($\mu$A/cm$^2$)}  & {$g_{K}$ (mS/cm$^2$)}  & {$R$ (k$\Omega$cm$^2$)}& extra connection\\ \hline
    $(19,56)_{RA}$& -10.95&0.17& 0   & 36  &0.00675&with $(19,19)_{LA}$ \\
    $(19,19)_{LA}$&  -10.95&0.17& 0       & 36   &0.00675&with $(19,56)_{RA}$\\
    $P_{\hbox{SAN}} =(19,19)_{RA}$& -10.95&0.17& 2.95   & 36  &0.00675& no \\
    Ventricle cells&  -10.95&0.17& 0       & 16.54   &0.00675& no \\
    All other cells&  -10.95&0.17& 0       & 36   &0.00675& no\\ \hline
\end{tabular}
}
\label{tab:snaps_wpw}
\end{table*}

\begin{figure}
    \centering
    \includegraphics[width=0.20\textwidth]{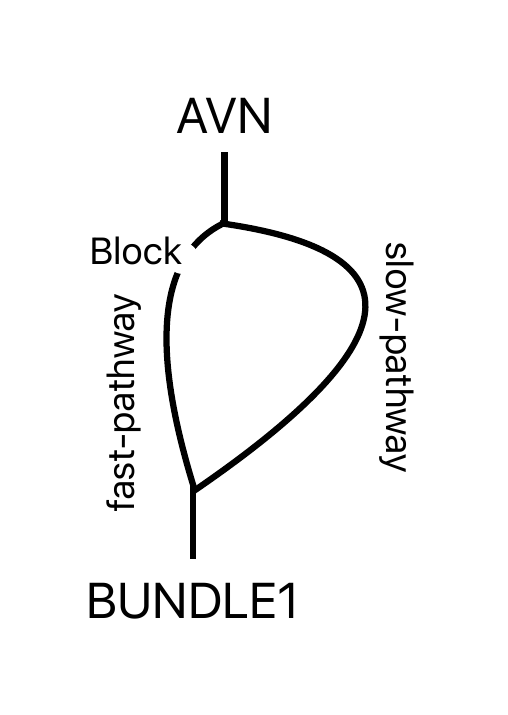} 
    \caption{Schematic view of the geometry of the Atrioventricular Node. The upper and lower points connect to both the AVN cell and BUNDLE1, respectively. The fast pathway shares the same resistivity as the other heart cells, while the slow pathway has a resistivity of $R_{slow} = 0.223$~ k$\Omega$ cm$^2$. The unidirectionally blocked cell is placed in the fast pathway.}
    \label{fig:avn_struc}
\end{figure}

 To model this mechanism, the slow pathway is modelled by reducing the diffusion coefficient, which does not contribute  To model this mechanism, the slow pathway is represented by decreasing the diffusion coefficient, which does not affect the shortened refractory period. However, by using a more artificial method—specifically, creating a unidirectional block in one of the cells within the slow branch—we can simulate this behaviour and induce reentry. In figure~\ref{fig:avn_struc}, we depict the AVN structure. 
to the reduced refractory period. Nonetheless, by employing a more artificial approach, specifically by creating a unidirectional block on one of the cells within the slow branch, we can simulate this behaviour and induce reentry. In figure~\ref{fig:avn_struc}, we illustrate the AVN structure.

 The slow pathway has a gap-junction resistivity that is 33 times higher, with $R_{slow} = 0.223$~k$\Omega$cm$^2$. The cell at $P=(6,2)_{AVN}$ (top left branch) exhibits a unidirectional block, indicating that diffusion is hindered. The top cell of the AVN connects to $P=(57,57)_{RA}$ (the original position of the AVN), while the bottom cell of the AVN links with the first cell of the bundle. Initial conditions are provided in Table~\ref{tab:avn_snaps}. 
 
\begin{figure}[h]
    \centering
    \includegraphics[width=0.5\textwidth]{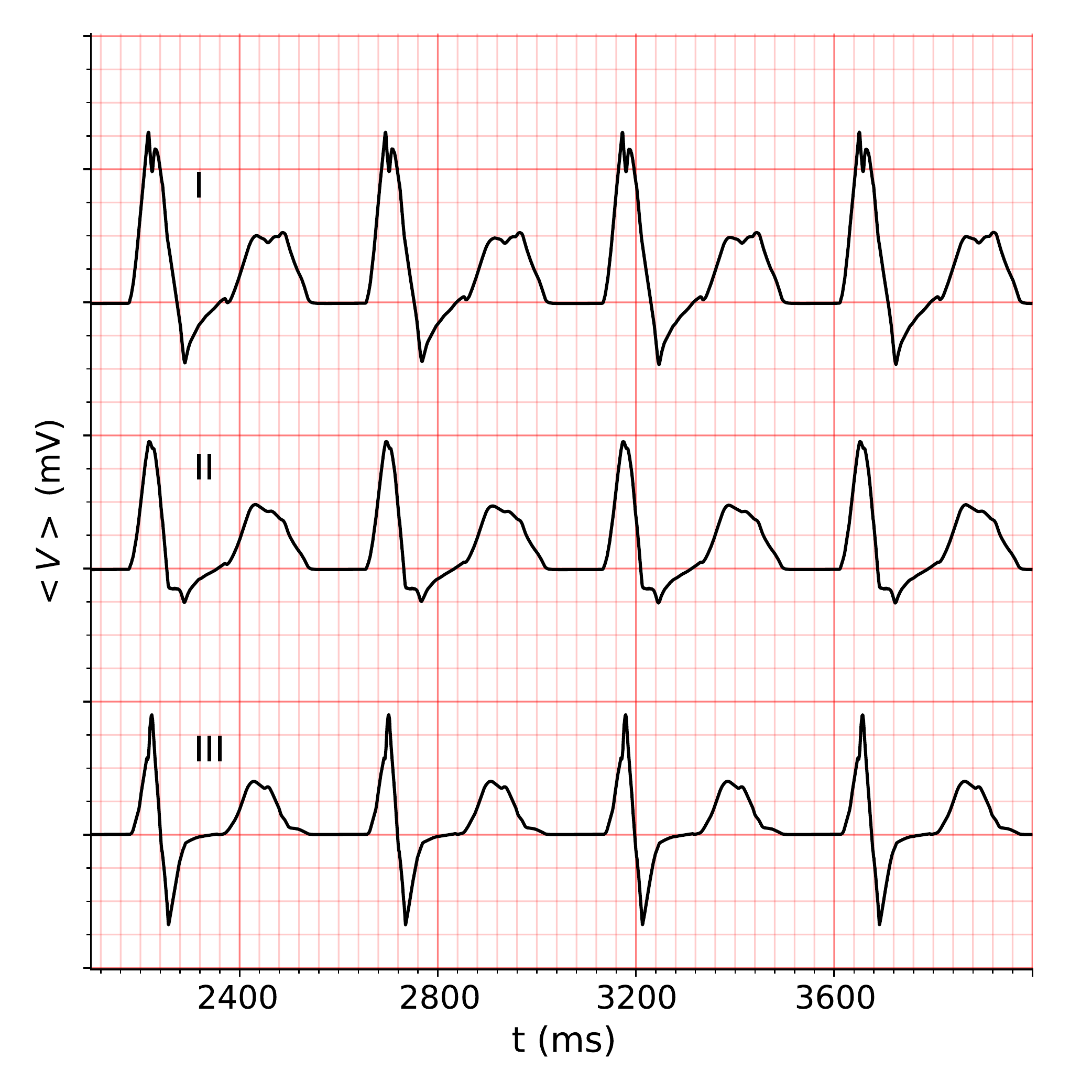} 
    \caption{Simulated the 3-lead ECG of the heart shows Atrioventricular Nodal Reentry Tachycardia, with the initial conditions presented in Table \ref{tab:avn_snaps}. The P-wave is only visible at the beginning ($t\simeq 400$~ms) before it is completely obscured by a repetitive sequence of QRS complexes followed by T-waves.}
    \label{fig:avn_ecg}
\end{figure}

 All of this is also illustrated in the 3-lead ECG shown in figure~\ref{fig:avn_ecg}. The standard three-wave ECG is replaced by a two-wave recording, with the P-wave obscured due to its coincidence with the QRS complex. This ECG pattern is a definitive sign that a patient is experiencing Atrioventricular Nodal Reentry Tachycardia. In this case, the heart rate for this tachycardia is 150 bpm, while the SAN is functioning at 60 bpm ($I_{SAN} = 2.95$~$\mu$A/cm$^2$).

\subsection{Accessory pathway}

\begin{figure}
    \centering
    \includegraphics[width=0.5\textwidth]{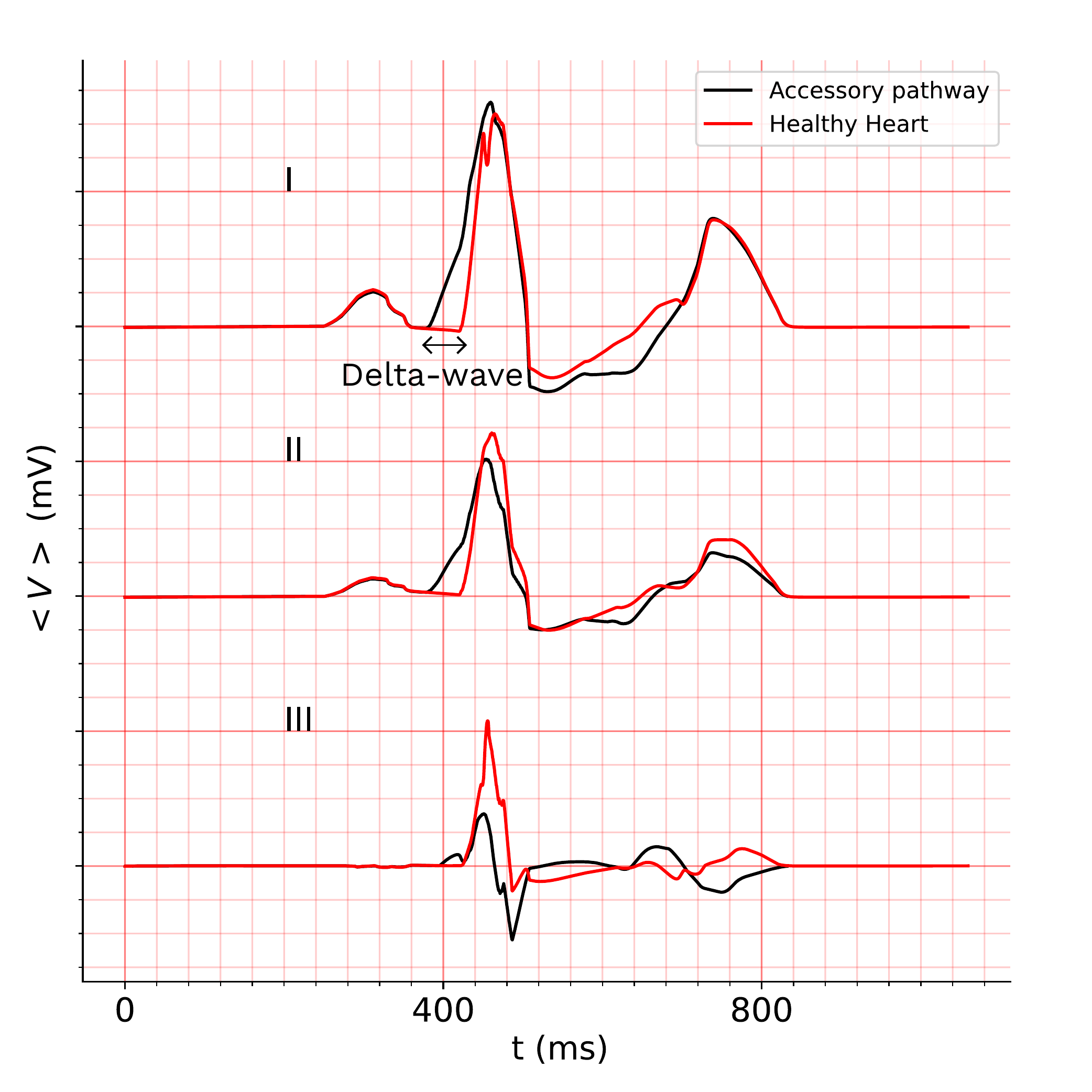} 
    \caption{In black, a simulated 3-lead ECG of the heart, featuring an accessory pathway that connects the RA and RV is displayed. The initial conditions are outlined in Table~\ref{tab:snaps_wpw}. In contrast, the three-lead ECG of a healthy heart is shown in red. The key difference between the two is the early QRS complex known as the Delta-wave.}
    \label{fig:ecg_wpw}
\end{figure}

An accessory pathway is an unwanted connection between areas of heart tissue that can cause problems with coordinating muscle contractions. Wolff-Parkinson-White syndrome (WPW), a congenital condition, exemplifies a scenario where excess tissue links two parts of the conduction system that should be electrically isolated 
\cite{Sap}. This syndrome is characterised by a connection between an atrium and a ventricle, resulting in the premature contraction of the latter. In this case, we focused on the connections on the right side of the human heart, specifically between the right atrium and the right ventricle. The initial conditions are detailed in Table~\ref{tab:snaps_wpw}, and the ECG results are depicted in Figure \ref{fig:ecg_wpw} in black, overlaid with the results from a healthy heart. Unsurprisingly, the QRS complex begins earlier than expected, leading to the delta wave, which is the primary feature of WPW syndrome.

\begin{table*}
\centering
\caption{Initial conditions and parameters for the model of a heart suffering from ischaemia on the right atrium.}
{\footnotesize
\begin{tabular}{cccccc} \hline
    {cells} & {$V(t = 0)$ (mV)} &{$n(t = 0)$}&{$I$ ($\mu$A/cm$^2$)}  & {$g_{K}$ (mS/cm$^2$)}  & {$R$ (k$\Omega$cm$^2$)}\\ \hline
    $(25\leq x\leq50,25\leq y\leq50)_{RA}$& -10.95&0.17& 0   & 36  & (0.00675,2.025] \\
    SAN $(19,19)_{RA}$& -10.95&0.17& 2.95   & 36  &0.00675 \\
    Ventricle cells&  -10.95&0.17& 0       & 16.54   &0.00675 \\
    All other cells&  -10.95&0.17& 0       & 36   &0.00675\\ \hline
\end{tabular}
}
\label{tab:diag_isch}
\end{table*}

\subsection{Ischemia-related arrythmias}

A lack of blood supply to myocytes can lead to their death and, more critically, to an incomplete apoptosis process that makes gap junctions semi-functional \cite{Yas}. Under these circumstances, the connections between adjacent cells become inefficient, resulting in a significant reduction in conduction velocities. These asymmetries in the muscular tissue can then cause wavefront separation and ultimately lead to re-entry, which causes arrhythmic events.

To model this condition, a square region $(25\leq x\leq50,25\leq y\leq50)$ was defined, where the resistivities of the gap junctions were randomly assigned between the standard $0.00675$~k$\Omega$cm$^2$ and an infinitely high resistivity, rendering the tissue dead and unresponsive. The random distribution was structured so that $80\%$ of the cells within this region were impaired, that is, with $R \in (0.00675, 2.025]$~k$\Omega$cm$^2$ and $15\%$ comprised of dead cells, while the remaining $5\%$ were completely healthy. Furthermore, the level of impairment among the $80\%$ is linearly related to the proximity to the centre of the scar: the closer a cell is to the centre of the square, the greater the probability that its gap junction resistivities will be elevated. Initial conditions are shown in Table~\ref{tab:diag_isch}.

\begin{figure}
    \centering
    \includegraphics[width=0.5\textwidth]{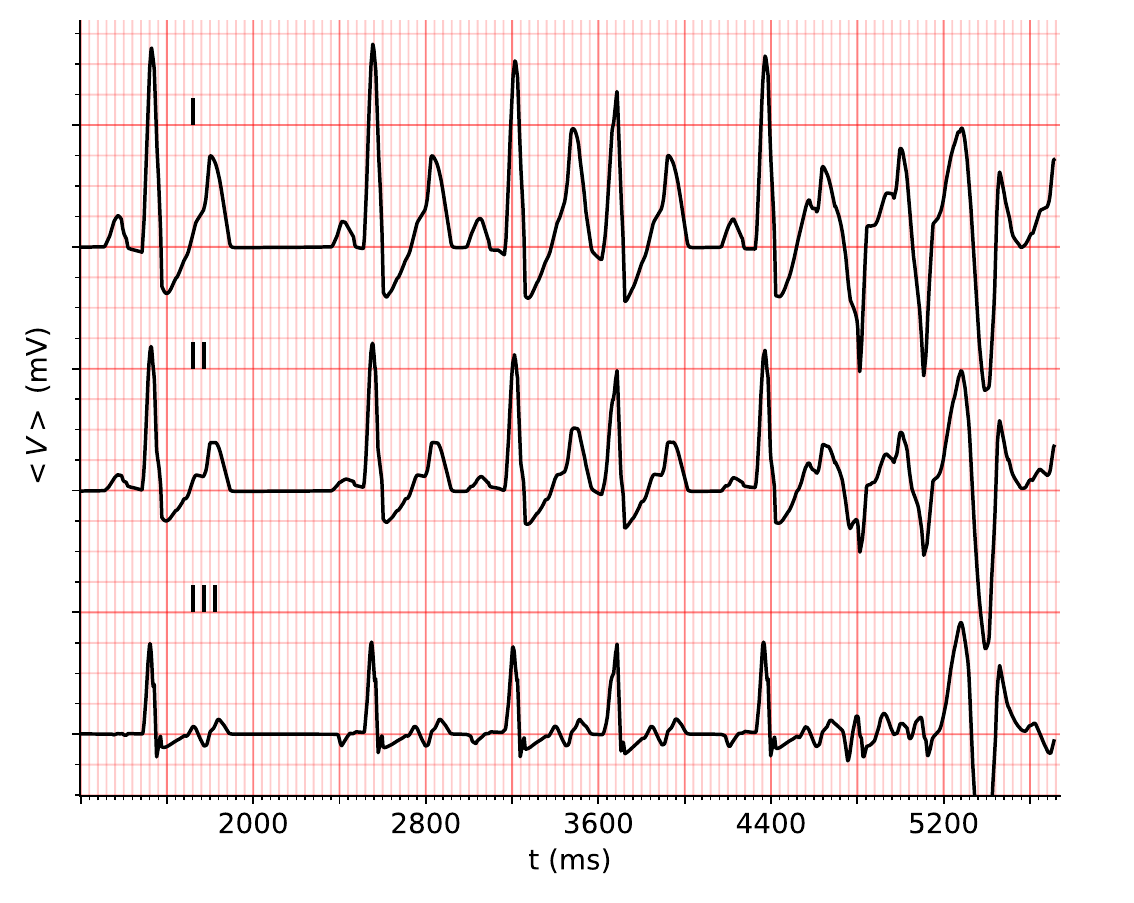} 
    \caption{Simulated 3-lead ECG of the model heart experiencing ischaemia at the RA, with initial conditions shown in Table \ref{tab:diag_isch}. The normal ECG waves are replaced by irregular electrical signatures at $t = 3040$~ms, due to re-entry triggered by the damaged tissue.}
    \label{fig:ecg_isch}
\end{figure}

Since this ischaemia is present in the RA, no significant changes in the overall shape of the ECG waves were expected, as this cavity contributes relatively little to the results of the leads. Aside from the increase in heart rate caused by reentrant waves within the damaged tissue, we expected only the P-wave to show a different shape, while both the QRS complex and the T-wave would stay similar to those of a healthy heart. However, as shown in figure~\ref{fig:ecg_isch}, this was not the case. Very irregular patterns appeared due to an issue with the excitation of the ventricles. Further work is needed to ensure the correct connection with the ventricles.

\section{Conclusions}
\label{sec:concl}

In this paper, we present a novel geometric whole-heart model that includes the atria and ventricles. Each of the four chambers is represented as a two-dimensional sphere where electrical signals propagate via diffusion. The connections between the atria are facilitated by the Bachmann bundle. Connections between the atria and ventricles are modelled by a one-dimensional cable representing the His and Purkinje fibres, as well as the atrioventricular node. Using this model, we have successfully simulated the electrical properties and repolarisation dynamics across the heart tissues. 

The electrical stimulus of the heart originates in the sinoatrial node through an action potential signal described by a simplified Hodgkin-Huxley type model, which features a single potassium-sodium gating variable. Adjustments to a current parameter of the sinoatrial node influence heart rate. The temporal evolution of action potentials across cardiac cells was simulated using Python. With this model, we generated an in silico 3-lead electrocardiogram. Simulations have shown that the 3-lead electrocardiogram closely matches clinical data from healthy individuals. These results suggest that this model can be regarded as a digital twin of the heart.  

Based on these results, we simulated several pathologies, developing the electrophysiological map of the heart along with the corresponding 3-lead electrocardiogram. Specifically, we modelled the conditions leading to arrhythmias, ventricular tachycardia, atrioventricular nodal reentry tachycardia, accessory pathways, and ischaemia-related arrhythmias. Our findings align with the commonly observed ECG waves for the various pathologies studied, demonstrating that this model can be a valuable tool in human heart research, providing ECG signatures that can validate results against clinically obtained data. Due to the relatively low processing power needed to run this model, it offers a practical way to analyse different pathological scenarios.

During the construction of the model, we faced several challenges related to the various connections between the heart chambers, the initial conditions that caused abnormal propagation properties in the cardiac tissues, the formation of spiral waves, and the effect of myocytes' ionic gating variables. Improving our understanding of these anatomical and physiological processes is a vital part of cardiac research.

\bigskip
\noindent \textbf{Acknowledgements:} We would like to thank S\'ergio Laranjo and Pedro Cunha, cardiologists from Santa Marta Hospital in Lisbon, for introducing us to the dynamics and physiology of the heart. This article has benefited from discussions with them and extensive reading. We also thank our colleague Teresa Pinheiro for her biological support and unwavering enthusiasm for this research.

\bigskip
\noindent \textbf{Conflict of interests:} The authors declare no competing interests.

\bigskip
\noindent \textbf{Author contribution:}  JO and RD have participated in the design of the model. JO contributed to software development, ECG and pathologies modelling, and validation of results. RD contributed to the conceptualisation and the geometric construction of the model. JO wrote the first draft of the model. Both authors read and approved the final manuscript.

\begin{multicols}{2}
  \null \vfill
  \includegraphics[width=0.2\textwidth]{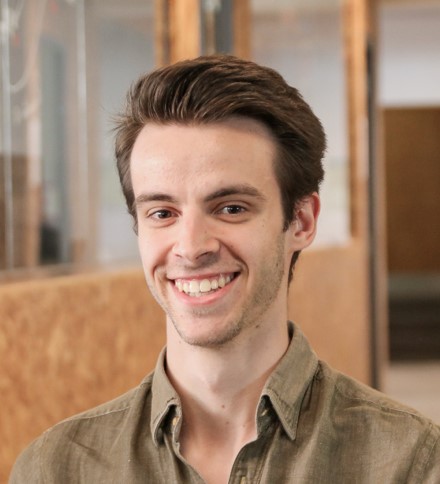}
  \vfill \null
\columnbreak
  \null \vfill
\noindent Jo\~ao Ol\'{\i}via is a PhD candidate and received a MSc in Physical Engineering. His research interests involve cardiac electrophysiology, data analysis and biophysical modelling.
  \vfill \null
\end{multicols}

\begin{multicols}{2}
  \null \vfill
  \includegraphics[width=.2\textwidth]{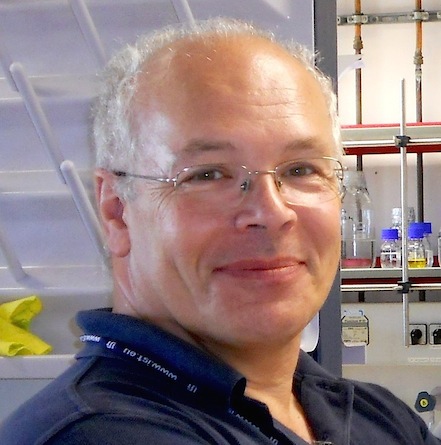}
  \vfill \null
\columnbreak
  \null \vfill
\noindent  Rui Dil\~ao is a Professor of Dynamical Systems and Biophysics at the University of Lisbon. His research interests focus on dynamical systems and biophysical modelling. 
  \vfill \null
\end{multicols}
\end{document}